\documentclass[summary]{gass2017}

\usepackage{cite, xcolor, threeparttable}

\title{Experimental Verification of Reflectionless Wide-Angle Refraction \\via a Bianisotropic Huygens' Metasurface}

\author{Michael Chen*\affref{ref1}, Elena Abdo-S\'{a}nchez\affref{ref1}\affref{ref2}, Ariel Epstein\affref{ref3},
  and George V. Eleftheriades\affref{ref1}}

\affiliation{%
  	\aff{ref1}{The Edward S. Rogers Sr. Department of Electrical and Computer Engineering,
University of Toronto,\\Toronto, ON, Canada M5S 2E4}
	\aff{ref2}{Dpto. Ingenier\'{i}a de Comunicaciones, E.T.S.I. Telecomunicaci\'{o}n, Universidad de M\'{a}laga,\\Andaluc\'{i}a Tech, E-29071 M\'{a}laga, Spain}
  	\aff{ref3}{Andrew and Erna Viterbi Faculty of Electrical Engineering,
Technion - Israel Institute of Technology, Haifa 3200003, Israel}
}

\begin{document}

\maketitle

\begin{abstract}
We report the design, fabrication, and characterization of bianisotropic Huygens' metasurfaces (BHMSs) for refraction of normally incident beams towards 71.8 degrees. As previously shown, all three \textcolor{black}{BHMS} degrees of freedom, namely, electric polarizability, magnetic polarizability and omega-type magnetoelectric coupling, are required to ensure no reflections occur for such wide-angle impedance mismatch. The unit cells are composed of three metallic layers, yielding a printed-circuit-board (PCB) structure. The fabricated BHMS is characterized in a quasi-optical setup, used to accurately assess specular reflections. Subsequently, the horn-illuminated BHMS' radiation pattern is measured in a far-field chamber, to evaluate the device's refraction characteristics. The measured results verify that the BHMS has negligible reflections, and the majority of the scattered power is coupled to the desirable Floquet-Bloch mode. To the best of our knowledge, this is the first experimental demonstration of such a reflectionless wide-angle refracting metasurface.
\end{abstract}

\section{Introduction}
\label{sec:introduction}
Engineered plane-wave refraction was one of the first functionalities demonstrated with a metasurface \cite{Yu2011}. These ultrathin planar devices are comprised of subwavelength polarizable particles (meta-atoms), allowing interaction with applied fields on a subwavelength scale via equivalent boundary conditions \cite{Holloway2012, Tretyakov2015}. It was soon found that to efficiently couple the incident beam towards a given direction in transmission, meta-atoms with both electric and magnetic polarizabilities have to be used \cite{Pfeiffer2013, Monticone2013, Selvanayagam2013}. Nonetheless, these so-called Huygens' metasurfaces (HMSs) feature a symmetric structure \cite{Monticone2013}, which only allows matching the wave impedance of either the incident or refracted waves \cite{Epstein2014_2}. Due to the inevitable mismatch, specular reflections occur, becoming significant for wide-angle refraction \cite{Selvanayagam2013}. 


This issue was solved in \cite{Wong2016}; by breaking the symmetry of HMS meta-atoms, a metasurface that is impedance matched for both incident and refracted fields was devised. This concept was later generalized in \cite{Epstein2016_3}, showing that \textcolor{black}{the required} asymmetric meta-atoms feature omega-type bianisotropy \textcolor{black}{(this was independently derived in \cite{Asadchy2016})}, exhibiting magnetoelectric coupling in addition to electric and magnetic polarizability. Correspondingly, it was shown that these \textcolor{black}{bianisotropic Huygens' metasurfaces (BHMSs)} can be realized by \emph{asymmetric} cascade of three impedance sheets \cite{Wong2016, Epstein2016_3}.

\begin{figure}[t]
\centering
\includegraphics[width=7.1cm]{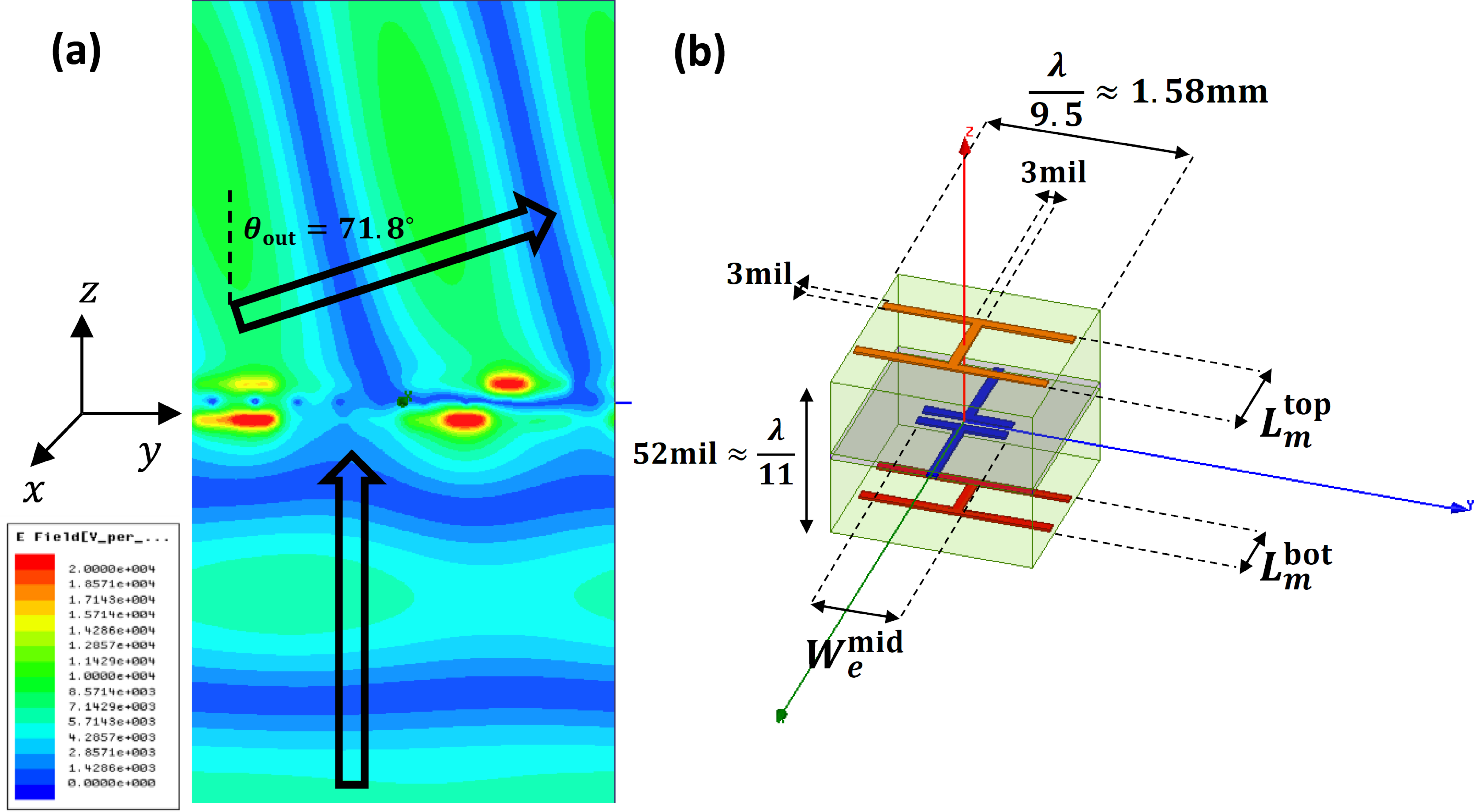}
\caption{(a) Refracting \textcolor{black}{BHMS} (one period) and simulated field distribution $\Re\left\{\left|E_x\left(y,z\right)\right|\right\}$ [a.u.]. (b) Meta-atom geometry and parameters.}
\label{fig:metasurface_configuration}
\end{figure}

In this paper, we verify \textcolor{black}{this theoretical concept}, by designing, fabricating, and characterizing a \textcolor{black}{BHMS} for reflectionless refraction of a transverse-electric normally-incident beam (${\theta_\mathrm{in}=0}$) towards ${\theta_\mathrm{out}=71.8^{\circ}}$ at ${f\sim20\mathrm{GHz}}$ [Fig. \ref{fig:metasurface_configuration}(a)]. The design \textcolor{black}{followed \cite{Epstein2016_3}}, yielding a printed circuit board (PCB) layout for the \textcolor{black}{BHMS}, verified via full-wave simulations. The fabricated PCB was then characterized in a quasi-optical setup, \textcolor{black}{accurately assessing} specular reflections. To complement this, the horn-illuminated \textcolor{black}{BHMS}' radiation pattern was measured in a far-field chamber, evaluating its refraction properties. The measured results indicate that, indeed, the specular reflections from the \textcolor{black}{BHMS} are negligible, and that the majority of the scattered power (${\sim80\%}$) is coupled to the desirable Floquet-Bloch (FB) mode, propagating towards $71.8^{\circ}$. To the best of our knowledge, this is the first experimental demonstration of such a reflectionless wide-angle refracting metasurface; it verifies the theory as well as demonstrates the viability of PCB \textcolor{black}{BHMS}s for realizing future omega-bianisotropic devices \cite{Asadchy2015,Epstein2016_3,Asadchy2016,Epstein2016_4}.

\section{Theory, Design, and Physical Realization}
\label{sec:theory_design}
The derivation in \cite{Epstein2016_3} formulates the spatially-dependent electric, magnetic, and magnetoelectric responses required for implementing the desirable functionality. In the specific case of plane-wave refraction, this response can be naturally expressed via generalized scattering matrix $\mathbf{[G]}$ parameters, with the port impedances corresponding to the wave impedances of the incident and refracted modes \cite{Wong2016, Epstein2016_3}. Within this framework, the \textcolor{black}{BHMS} should be composed of meta-atoms at positions $y$, which exhibit unity (generalized) transmission coefficients ${\left|G_{21}\right|=1}$ and linear (generalized) transmission phase ${\angle G_{21}\left(y\right)=-\frac{2\pi}{\lambda}y\Delta_\mathrm{sin}+\xi_\mathrm{out}}$, where $\lambda$ is the wavelength, $\xi_\mathrm{out}$ is a constant phase, and ${\Delta_\mathrm{sin}=\sin\theta_\mathrm{out}-\sin\theta_\mathrm{in}}$. This guarantees perfect impedance matching at both ports, while providing the necessary change in the transverse wavenumber to fully-couple the incident wave to the refracted one. 

To implement the \textcolor{black}{BHMS}, the cascaded impedance sheet scheme of \cite{Wong2016, Epstein2016_3} was used. Specifically, the \textcolor{black}{BHMS} consists of three copper layers ($1/2\mathrm{oz.}$), defined on two $25\mathrm{mil}$ Rogers RT/duroid 6010 laminates; the latter are bonded using a $2\mathrm{mil}$ Rogers 2929 bondply, yielding an overall metasurface thickness of ${52\mathrm{mil}\approx\lambda/11}$ at the design frequency ${f=20\mathrm{GHz}}$. Every meta-atom has ${\Delta_x\times\Delta_y=\lambda/9.5\times\lambda/9.5\approx1.58\mathrm{mm}\times1.58\mathrm{mm}}$ lateral dimensions, such that the metasurface period ${\lambda/\Delta_\mathrm{sin}}$ contains 10 unit cells. Each unit cell consists of a dogbone, a loaded dipole, and another dogbone, forming the bottom, middle, and top impedance sheets, respectively [Fig. \ref{fig:metasurface_configuration}(b)]. The meta-atom response is controlled by the dogbone arm lengths, $L_m^\mathrm{bot}$ and $L_m^\mathrm{top}$, and the capacitor width $W_e^\mathrm{mid}$. These parameters were initially set for each meta-atom following analytical formulas relating the required \textcolor{black}{BHMS} parameters to the sheet impedances, and fine tuned via simulations (ANSYS HFSS) to achieve the required overall bianisotropic response \cite{Epstein2016_3}.

\begin{figure}
\centering
\includegraphics[width=7.5cm]{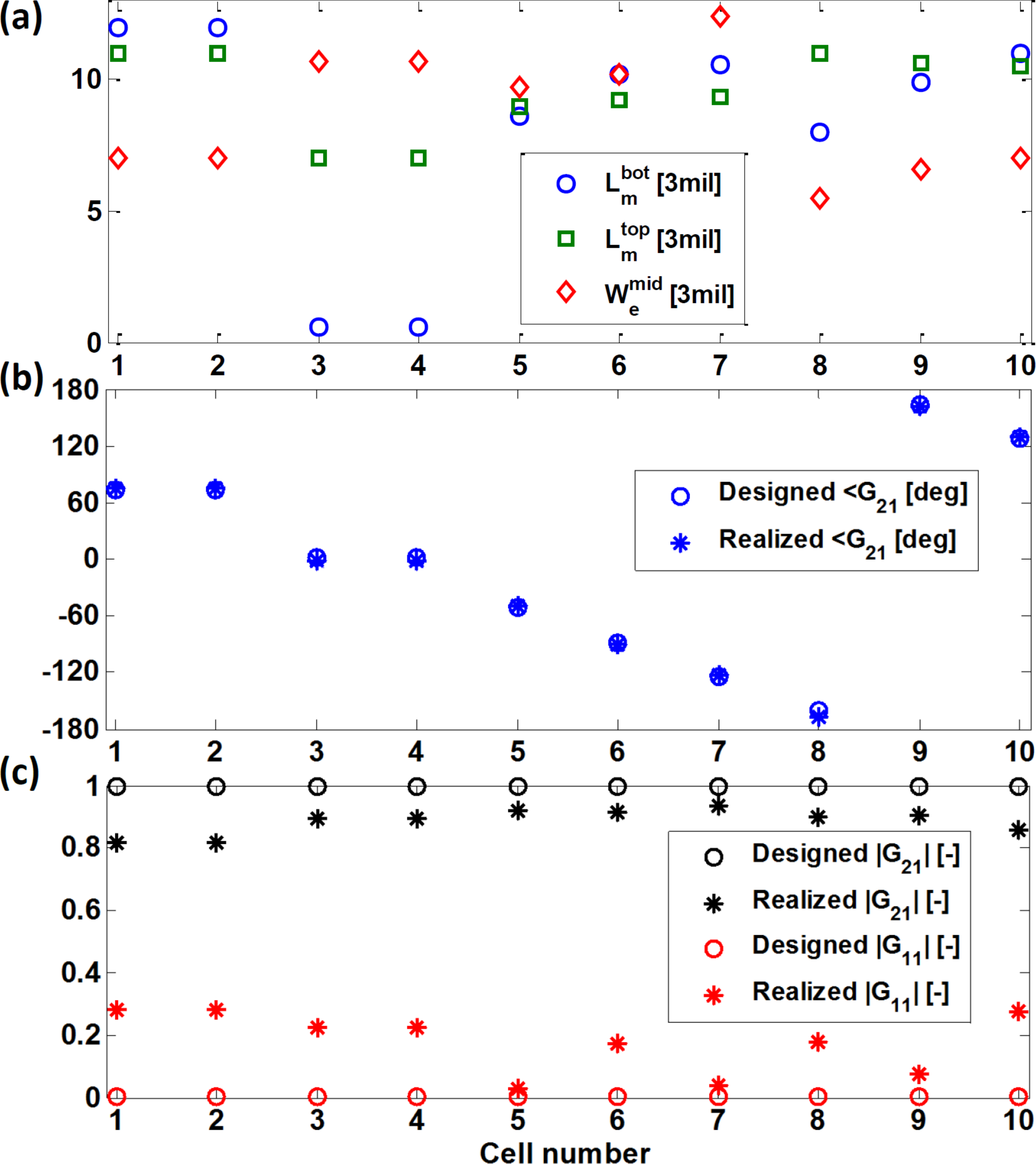}
\caption{\textcolor{black}{BHMS} design. (a) Optimized geometrical parameters of the meta-atoms [corresponding to Fig. \ref{fig:metasurface_configuration}(b)]. (b) Designed ($\circ$) and realized ($\ast$) generalized transmission phase. (c) Designed ($\circ$) and realized ($\ast$) generalized transmission (black) and reflection (red) magnitudes.}
\label{fig:metasurface_design}
\end{figure}

The final element geometrical parameters are presented in Fig. \ref{fig:metasurface_design}(a), along with the desirable and achievable \textcolor{black}{responses} [Fig. \ref{fig:metasurface_design}(b)-(c)], which generally agree well (losses cause reduction in $\left|G_{21}\right|$). 
Due to losses exhibited by the meta-atoms with $\angle G_{21}$ around $0^\circ$, we used at two occasions identical unit cells \textcolor{black}{(cells \{\#1,\#2\} and \{\#3,\#4\})} that implement the average response of two consecutive elements, which improved the overall transmission. 

One period of the metasurface was simulated under periodic boundary conditions, yielding the field distribution presented in Fig. \ref{fig:metasurface_configuration}(a). Simulations predict that about $28\%$ of the incident power is absorbed in the metasurface. Out of the scattered power, $93\%$ is coupled to the desirable FB mode (transmitted towards $71.8^\circ$), $3\%$ is specularly reflected, and $4\%$ is transmitted towards $-71.8^\circ$. \textcolor{black}{A frequency scan reveals that specular reflections are minimal at $19.8\mathrm{GHz}$; however, the main device characteristics, i.e. the refraction efficiency and ohmic losses, are similar within the range ${19.8-20.0\mathrm{GHz}}$}. This verifies that the designed \textcolor{black}{BHMS} indeed implements reflectionless wide-angle refraction as prescribed. The relatively high losses probably originate in the resonant nature of the meta-atoms, and could be improved by \textcolor{black}{additional} impedance sheets \cite{Pfeiffer2014_3}.

Following the verified design, a metasurface PCB of size ${12''\times18''=30.48\mathrm{cm}\times45.72\mathrm{cm}\approx20\lambda\times30\lambda}$ was fabricated (Fig. \ref{fig:quasi_optical_setup}), containing ${29\times190}$ replicas of the simulated \textcolor{black}{BHMS} period [Fig. \ref{fig:metasurface_configuration}(a)] along the $y$ and $x$ axes, respectively. 


\section{Experimental Results}
\label{sec:results}

\subsection{Quasi-optical Specular Reflection Experiment}
\label{subsec:quasi_optical}
\begin{figure}
\centering
\includegraphics[width=7.8cm]{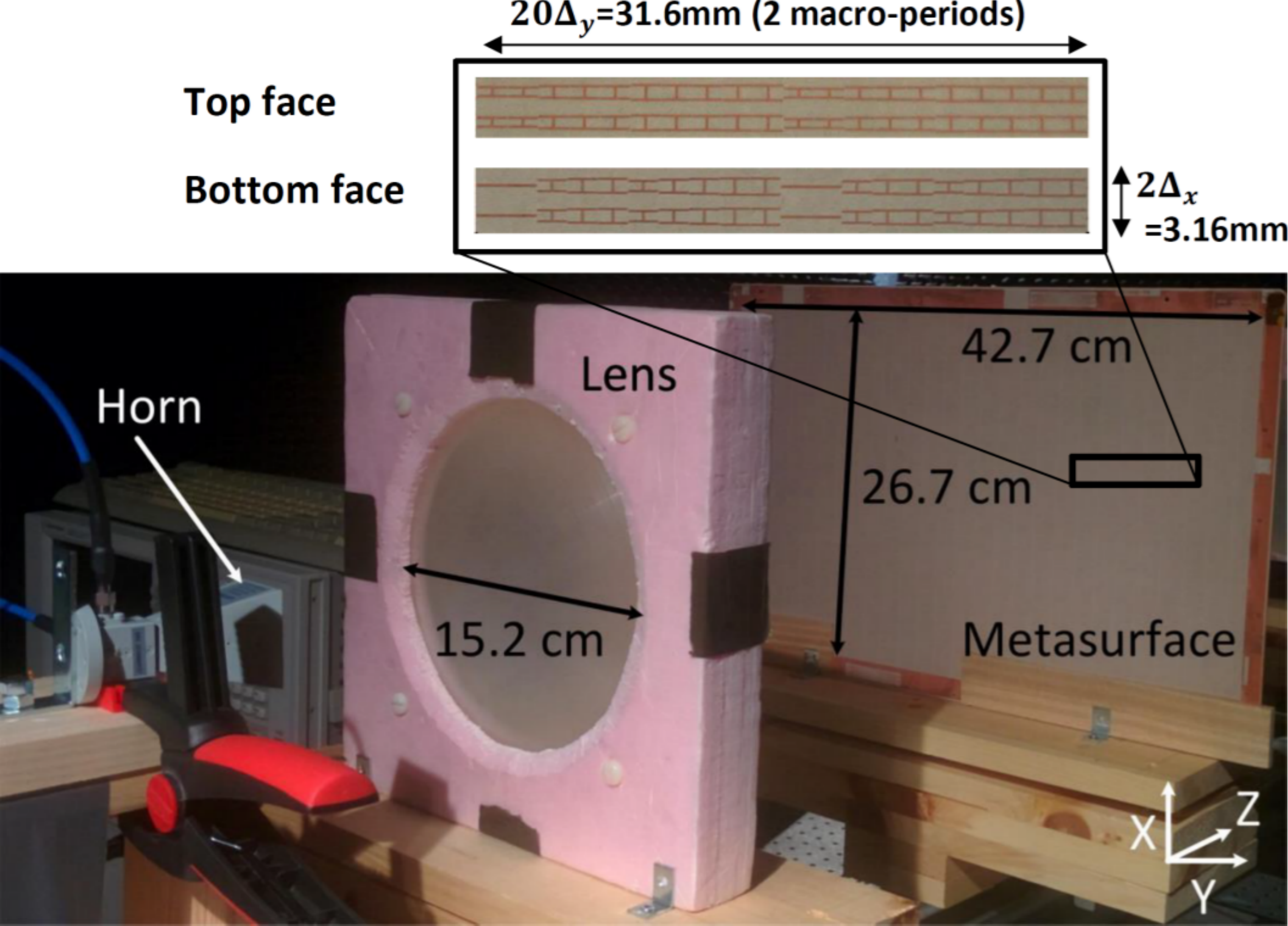}
\caption{Quasi-optical experimental setup. The focal distances from the lens to the horn and from the lens to the \textcolor{black}{BHMS} are 12.5 cm and 29 cm, respectively. \textbf{Inset}: Close-up on the top and bottom faces of the fabricated metasurface.}
\label{fig:quasi_optical_setup}
\end{figure}

To test the specular reflectionless nature of the metasurface, a quasi-optical experiment was designed. This setup uses a horn (A-info LB-OMT-150220) and a Rexolite lens to focus a Gaussian beam onto the \textcolor{black}{BHMS} (Fig. \ref{fig:quasi_optical_setup}). The horn and the \textcolor{black}{BHMS} are placed at the focal planes of the lens; thus, the wavefront incident upon the \textcolor{black}{BHMS} is planar, allowing the characterization to closely resemble theory and simulations. 

\begin{figure}
\centering
\includegraphics[width=7.8cm]{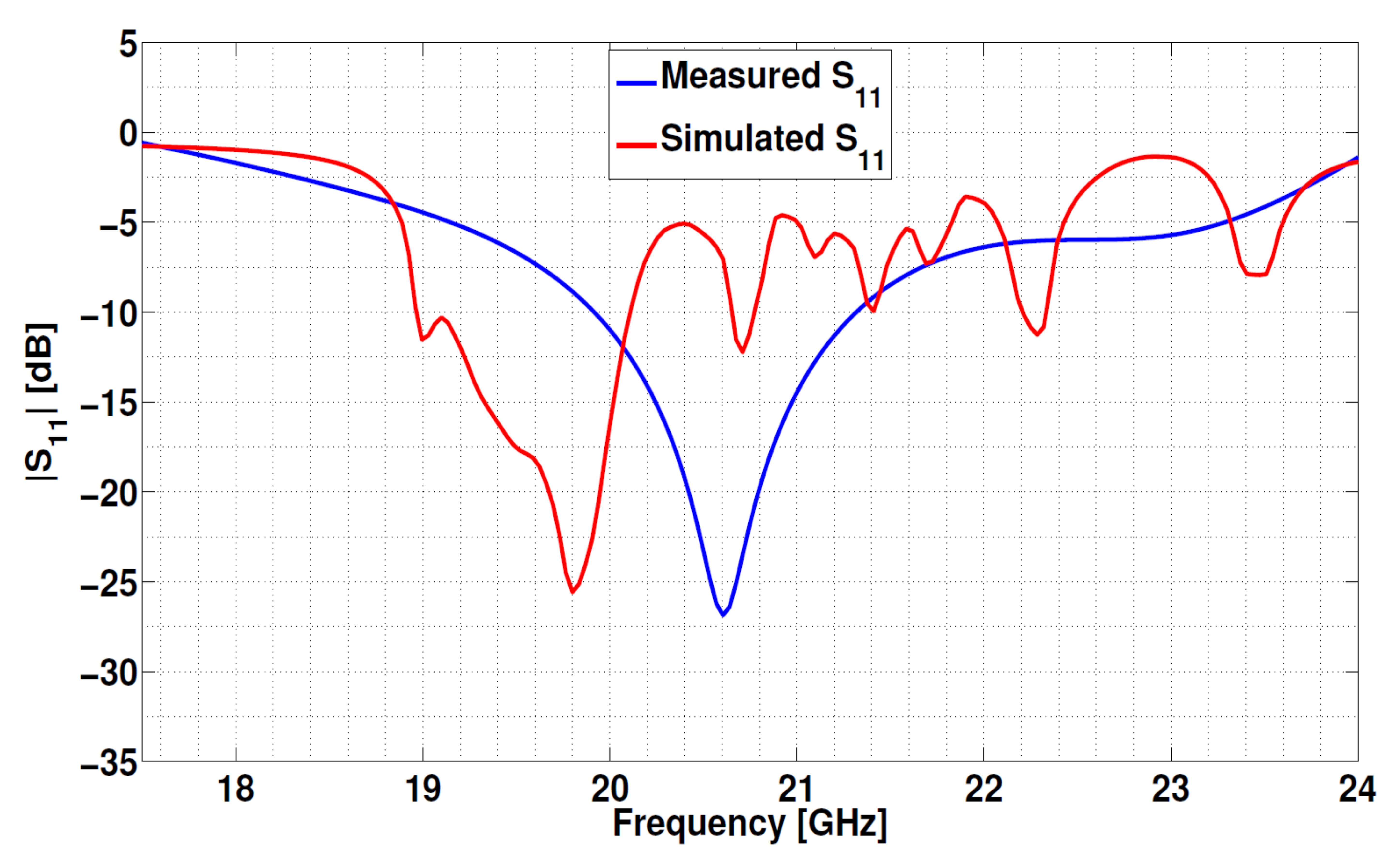}
\caption{Measured and simulated normally incident specular reflection.}
\label{fig:S11_response}
\end{figure}

The measured and simulated \textcolor{black}{$\left|S_{11}\right|$} versus frequency are presented in Fig. \ref{fig:S11_response}, revealing a shift of the resonant frequency from \textcolor{black}{$19.8\mathrm{GHz}$ (simulated) to $20.6\mathrm{GHz}$ (measured), which is attributed to fabrication errors and material uncertainties. The measured S$_{11}$ indicates that less than 0.2\% of the incident power is back-reflected, in agreement with simulations, verifying that the \textcolor{black}{BHMS} is indeed specularly reflectionless.}

\subsection{Far-field Anechoic Chamber Refraction Experiment}
\label{subsec:far_field}
As the main feature of the metasurface is its ability for extreme refraction, an experiment was conducted to characterize this effect. To this end, the \textcolor{black}{BHMS} was placed in front of a standard gain horn antenna (Quinstar QWH-KPRS-00)\textcolor{red}{,} and the radiation pattern of the overall system (horn+\textcolor{black}{BHMS}) was measured as a combined antenna under test (AUT).
The transmitting horn was placed sufficiently far from the \textcolor{black}{BHMS} to produce a planar wavefront on its ${z\rightarrow0^+}$ face, while the receiving horn was aligned behind the \textcolor{black}{BHMS}, facing its ${z\rightarrow0^-}$ face \textcolor{black}{(see coordinate system in Fig. \ref{fig:metasurface_configuration})}. 
Absorbers were attached to the back and \textcolor{black}{sides} of the receiving horn to reduce spurious scattering. 
The AUT was then rotated around its axis, and the gain pattern was measured, expected to yield a maximum around ${\theta_\mathrm{out}=71.8^\circ}$. 

We note that this test is not ideal and there are clear trade-offs in the experimental setup. 
One such parameter is the distance of the receiving horn from the \textcolor{black}{BHMS}. As the \textcolor{black}{BHMS} was designed to interact with plane waves, the horn should be placed sufficiently far away to match the expected planar wavefront of the refracted wave. However, since the \textcolor{black}{BHMS} has a finite size, if the horn is placed too far, it would not be sufficiently shadowed by the surface. 
A good compromise was found at a horn-\textcolor{black}{BHMS} separation distance of ${24\mathrm{cm}=16\lambda}$.

\begin{figure}
\centering
\includegraphics[width=7.8cm]{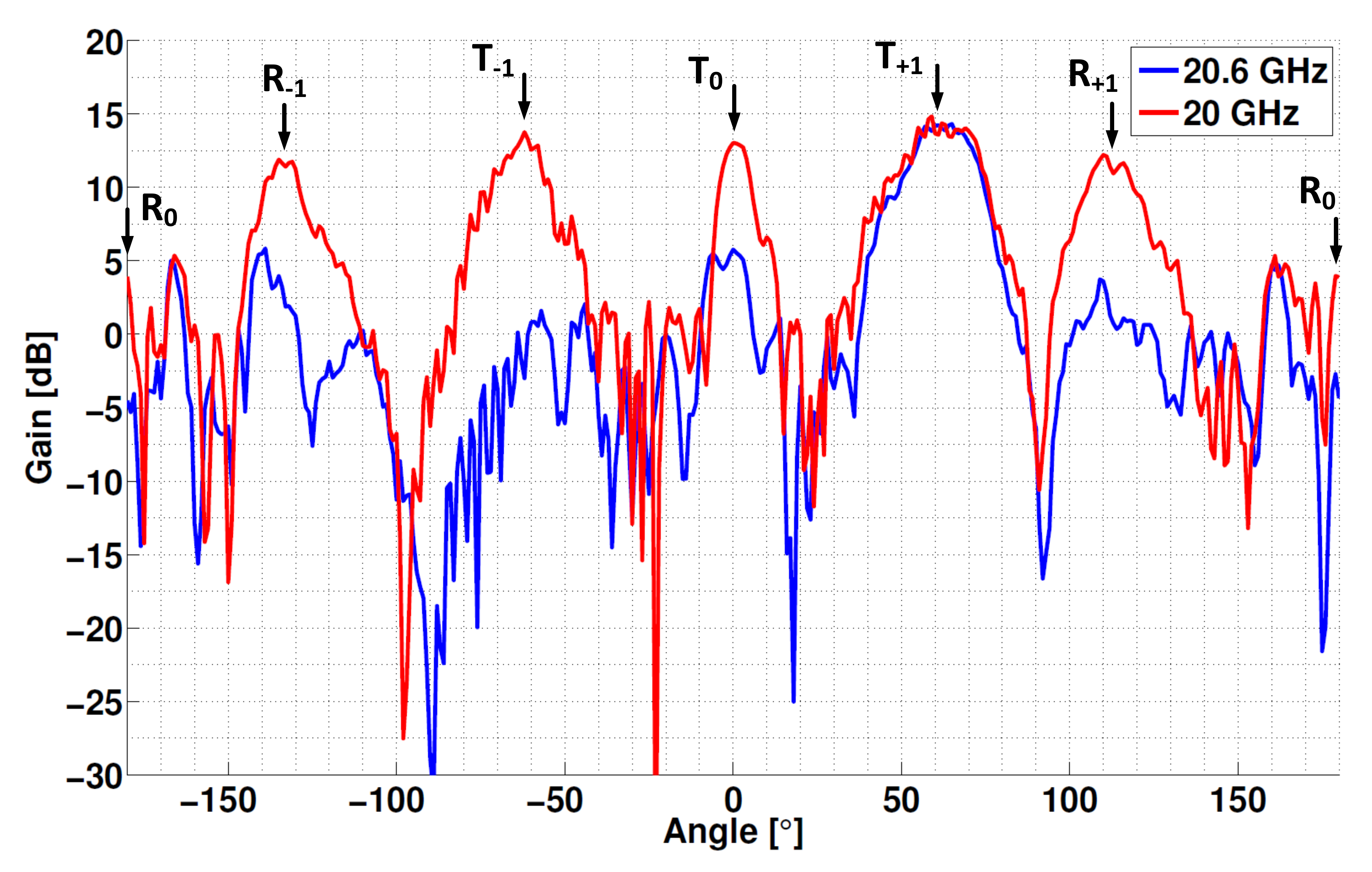}
\caption{Measured AUT radiation patterns at 20.6 GHz (resonant) and 20 GHz (off resonance). The excited FB modes are clearly identified: the 0th transmitted mode (0$^\circ$), the $\pm1$ transmitted modes ($\pm71.8^\circ$), the 0th reflected mode ($\pm180^\circ$) and the $\pm1$ reflected modes ($\pm108.2^\circ$).}
\label{fig:radiation_pattern}
\end{figure}

The measured radiation patterns at $20\mathrm{GHz}$ and $20.6\mathrm{GHz}$ are presented in Fig. \ref{fig:radiation_pattern}. While the entire measured angular range is shown, the most reliable information is obtained between -150$^\circ$ and 150$^\circ$. 
When the AUT is measured close to $\pm180^\circ$, the receiving horn is partially blocking the line of sight between the transmitter and the metasurface; as the back of the horn has been fitted with absorbers, this would interfere with the measured results. Nonetheless, the quasi-optical measurements complement the radiation pattern for these angles, assessing the specular reflection (Section \ref{subsec:quasi_optical}). 

The radiation patterns in Fig. \ref{fig:radiation_pattern} clearly indicate that the \textcolor{black}{BHMS} implements the desirable refraction functionality. At 20 GHz, away from the \textcolor{black}{(measured)} resonant frequency, all propagating FB modes are excited, corresponding to multiple scattered beams. However, at \textcolor{black}{resonance ($20.6\mathrm{GHz}$)}, most of the scattered power is coupled to the designated mode, refracting towards $\approx71.8^\circ$, while the other modes are suppressed. 

A closer look at the pattern at 20.6 GHz reveals that the gain actually peaks at 62$^\circ$. However, as the gain of finite apertures deteriorates with a cos($\theta$) factor, we must account for it to properly assess the direction of the main beam. This evaluates the refraction angle at 69$^\circ$, much closer to the designated one; the deviation may originate in fabrication tolerances and alignment errors. The 3dB beamwidth is $21^\circ$, corresponding to an effective aperture length of $20\mathrm{cm}$. Given the receiving horn dimensions and position, the expected $-10\mathrm{dB}$ beam diameter on the \textcolor{black}{BHMS} is ${\approx\!\!23\mathrm{cm}}$ \cite{Goldsmith1998_Chap7_GaussianHorn}, in a good agreement. 

Comparing the power in the refracted beam, calculated by integrating the gain between its nulls, to the overall integrated gain, indicates that approximately $80\%$ of the \emph{scattered} power is coupled to the desirable FB mode. Although this is smaller than the simulated $93\%$, which may be attributed to fabrication and material tolerances, it still demonstrates a reasonable quantitative agreement with the designated device functionality. It should be noted that these efficiencies, corresponding to \emph{realistic} \textcolor{black}{BHMS}s, are considerably higher than the theoretical $73\%$ predicted for an \emph{ideal} \textcolor{black}{non-bianisotropic} HMS implementing the same wide-angle refraction\cite{Selvanayagam2013,Epstein2014_2}; this highlights the crucial role omega-type bianisotropy plays in face of a significant impedance mismatch. 

\textcolor{black}{Figure \ref{fig:radiation_pattern} further indicates that the suppressed scattering at $20.6\mathrm{GHz}$ is mostly an outcome of increased absorption (the peak gain remains similar). Although this implies that losses become much pronounced at the resonant frequency, the current measurement setup does not allow reliable quantification of this parameter; this is left for future work. }


\section{Conclusion}
\label{sec:conclusion}
We have demonstrated the first PCB \textcolor{black}{BHMS} implementing reflectionless wide-angle refraction. Experimental validation was carried out by combining the results of a quasi-optical setup and radiation pattern measurements. The hybrid approach verifies that, indeed, specular reflections are negligible, and approximately $80\%$ of the \emph{scattered} power is coupled to the designated beam, albeit with considerable losses. Future work will include \textcolor{black}{a more thorough} quantification of losses, and using \textcolor{black}{improved} \textcolor{black}{BHMS} designs to \textcolor{black}{further} enhance the overall efficiency.

%


%
%

\end{document}